\documentclass[aps,amsmath,amssymb,reprint,]{revtex4-2}
\usepackage{graphicx}
\usepackage{dcolumn}
\usepackage{bm}
\usepackage[utf8]{inputenc}
\usepackage[T1]{fontenc}
\usepackage{mathptmx}
\usepackage{etoolbox}
\usepackage{array,multirow}

\usepackage{chemformula}
\makeatletter
\def\@email#1#2{%
	\endgroup
	\patchcmd{\titleblock@produce}
	{\frontmatter@RRAPformat}
	{\frontmatter@RRAPformat{\produce@RRAP{*#1\href{mailto:#2}{#2}}}\frontmatter@RRAPformat}
	{}{}
}%
\makeatother
\begin{document}
	
	\title{Large trion binding energy in monolayer WS$_2$ via strain-enhanced electron-phonon coupling}
	\author{Yunus Waheed$^1$\footnote[0]{$^{\text{*}}$Electronic mail: skumar@iitgoa.ac.in}}
	\author{Sumitra Shit$^{1}$}
	\author{Jithin T Surendran$^{1}$}
	\author{Indrajeet D Prasad$^{1}$}
	\author{Kenji Watanabe$^{2}$}
	\author{Takashi Taniguchi$^{3}$}
	\author{Santosh Kumar$^{1}$}
	\affiliation{$^1$School of Physical Sciences, Indian Institute of Technology Goa, Ponda 403401, Goa, India}
	\affiliation{$^2$Research Center for Electronic and Optical Materials, National Institute for Materials Science, 1-1 Namiki, Tsukuba 305-0044, Japan}
	\affiliation{$^3$Research Center for Materials Nanoarchitectonics, National Institute for Materials Science, 1-1 Namiki, Tsukuba 305-0044, Japan}

	\date{\today}

	\begin{abstract}
	Transition metal dichalcogenides and related layered materials in their monolayer and a few layers thicknesses regime provide a promising optoelectronic platform for exploring the excitonic- and many-body physics. Strain engineering has emerged as a potent technique for tuning the excitonic properties favorable for exciton-based devices. We have investigated the effects of nanoparticle-induced local strain on the optical properties of exciton, $X^0$, and trion, $X^\text{-}$, in monolayer WS$_2$. Biaxial tensile strain up to 2.0\,\% was quantified and verified by monitoring the changes in three prominent Raman modes of WS$_2$: E${^1_{2g}}$($\Gamma$), A$_{1g}$, and 2LA(M). We obtained a remarkable increase of 34\,meV in $X^\text{-}$ binding energy with an average tuning rate of 17.5\,$\pm$\,2.5\,meV/\% strain across all the samples irrespective of the surrounding dielectric environment of monolayer WS$_2$ and the sample preparation conditions. At the highest tensile strain of $\approx$2\%, we have achieved the largest binding energy $\approx$100\,meV for $X^\text{-}$, leading to its enhanced emission intensity and thermal stability. By investigating strain-induced linewidth broadening and deformation potentials of both $X^0$ and $X^\text{-}$ emission, we elucidate that the increase in $X^\text{-}$ binding energy is due to strain-enhanced electron-phonon coupling. This work holds relevance for future $X^\text{-}$-based nano-opto-electro-mechanical systems and devices.		
	\end{abstract}
	
	\maketitle
	Transition Metal Dichalcogenides (TMDs) are gaining significant interest in the scientific community due to possession of their unique optical and optoelectronic properties \textcolor{blue}{\cite{mak10,butler2013progress,mak2013tightly}}. The properties on account of which TMDs stand out from other 2D material platforms are: indirect to direct bandgap conversion when they are thinned down to monolayer (ML) thicknesses \textcolor{blue}{\cite{manzeli20172d,chhowalla2013chemistry,zheng2018light}}, leading to stronger photoluminescence (PL) emission in visible and near-infrared spectral ranges \textcolor{blue}{\cite{mak10,splendiani10}}, reduced dielectric screening resulting in enhanced Coulombic interactions in the ML regime gives rise to higher exciton binding energies, $E_{b, \,X^0}$ (in the range of 0.2\,-\,1.0 eV) \textcolor{blue}{\cite{mueller18,zhu15}} and strong light-matter interactions \textcolor{blue}{\cite{wurstbauer17}}. TMDs have proven to be an ideal candidate for the next generation digital electronics due to their considerable bandgap \textcolor{blue}{\cite{butler2013progress}}. Owing to their tunable bandgap, TMDs offer a platform for various electronic and opto-electronic applications \textcolor{blue}{\cite{wang2012electronics}}. To date, different methods like temperature, electric field, magnetic field, doping and strain have been employed to effectively tune the electronic and optoelectronic properties of TMDs \textcolor{blue}{\cite{mas20112d,wang2015strain,sasaki2016growth}}.
	
	\par MLs of widely investigated TMDs like MoS$_2$, WS$_2$, MoSe$_2$, WSe$_2$ show prominent emission peaks in their PL spectra that are redshifted by 15 to 41 meV \textcolor{blue}{\cite{zhu15,he2016strain,dastidar2024optically}} from the ground-state (1s) neutral-excitonic, $X^{0}$, emission peaks. The formation of a singly-negative-charged trion, $X^\text{-}$ due to intrinsic or substrate-related $n$-doping of the ML is the cause of such redshifted peaks. The energetic difference, $E_{b, \,X^-}\,=\,E_{X^-}\,-\,E_{X^0}$, where $E_{X^-}\left(E_{X^0}\right)$ is the emission energy of $X^\text{-} \left(X^{0}\right)$ transition, is referred to the binding energy (B.E.) of $X^\text{-}$ with $X^0$. The values of $E_{b, \,X^-}$ in ML-WS$_2$ falls in the range of 20\,-\,41\,meV \textcolor{blue}{\cite{mak2013tightly,ross2013electrical,mitioglu2013optical,michail2023tuning,he2016strain}} owing to the nature of different dielectric environments. A lower $E_{b, \,X^-}$ leads to thermal quenching due to lower thermal stability, resulting in the shortening of $X^\text{-}$ lifetimes \textcolor{blue}{\cite{golovynskyi2021trion,golovynskyi2020exciton}}. Thus, a requirement of increasing the $E_{b, \,X^-}$ arise for efficiently controlling the $X^\text{-}$.
	
	\par It has been demonstrated that the ramping up of excitation laser power results in an increase in $E_{b, ,X^-}$ by a maximum of 5 meV \textcolor{blue}{\cite{mitioglu2013optical,kesarwani2022control,chowdhury2021modulation}}. Additionally, the use of laser light with high-order orbital angular momentum and at a higher excitation power has shown a substantial decrease of $E_{b, \,X^-}$ by a maximum of 9\,meV \textcolor{blue}{\cite{kesarwani2022control}}. A limitation of this method is that the excitation power above a certain limit would lead to a local heating effect; hence, tuning over a wider range is not possible. Strain-tuning has emerged as a powerful technique to tune the electronic properties and lattice dynamics of ML and a few-layer TMDs. An introduction of strain into the TMDs layers modifies the bandgaps and alters the effective masses and mobilities of charge carriers \textcolor{blue}{\cite{amin2014strain}}, leading to increased device performances. Uniaxial strain upto $\approx$ 2.2\% \textcolor{blue}{\cite{conley2013bandgap}} and biaxial strain upto $\approx$ 2.5\% \textcolor{blue}{\cite{lloyd2016band}} have been introduced in 2D materials by bending the flexible substrates containing the flakes of 2D materials \textcolor{blue}{\cite{conley2013bandgap,chang2013high,park2018transparent,carrascoso2022biaxial,carrascoso2023improved,michail2020biaxial,michail2023tuning}}, by nano-indentation of 2D materials using sharp-tips like AFM-tip \textcolor{blue}{\cite{harats2020dynamics}} and tapered fiber-tip \textcolor{blue}{\cite{gelly2022probing}}, by pressurizing the cylindrical-cavity covered with the flakes \textcolor{blue}{\cite{lloyd2016band}}, and by creating a thermal expansion coefficient mismatch between two materials \textcolor{blue}{\cite{frisenda2017biaxial,plechinger2015control}}. Very recently, Henríquez-Guerra et al., see Ref.\,[\textcolor{blue}{\onlinecite{henriquez2023large}}], demonstrated the effects of biaxial strain on the $E_{b, \,X^-}$ in ML-WS$_2$ and have shown an increase in $E_{b, \,X^-}$ by $\approx $ 3 meV under the introduction of a 1.5\% compressive strain.
	\begin{figure}[htb]
		\includegraphics{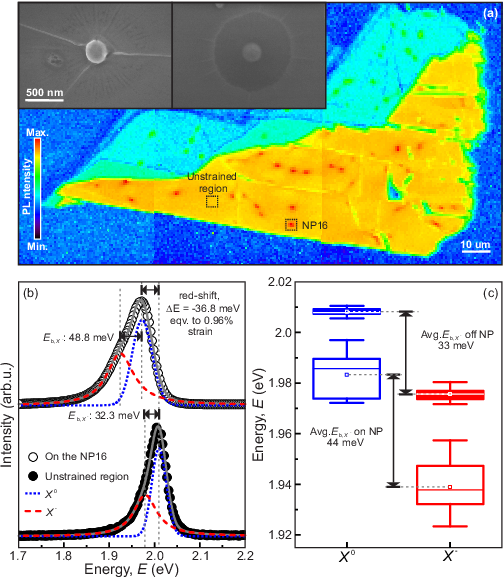}
		\caption{\textbf{\textmu-PL and SEM characterization of Sample\,1a: }(a) Combined image of color-coded PL peak intensity maps of Sample\,1a showing emission in the wavelength range of 535-880 nm. Inset: SEM images of two NPs with different flake conformalities : Conformality of flake according to the NP surface (left) and formation of tent-like structure on NP location. (b) Comparison of µ-PL spectra taken on-the-NP location (open circles, top) and unstrained region (closed circles, bottom) in ML-WS$_2$. Solid thick lines are the fits for the spectrum on-the-NP and unstrained locations. (c) Box-chart showing the $X^0$ and $X^\text{-}$ emission energies distribution for unstrained and strained sites in sample. Dotted gray lines show the average of the distribution for both strained and unstrained sites.}
		\label{Fig:strain1}
	\end{figure}
	\par  Here, we have investigated the effects of local biaxial tensile-strain on PL emission properties of $X^0$ and $X^\text{-}$ in ML-WS$_2$, and have demonstrated a significant increase in $E_{b, \,X^-}$ by 34\,meV and a 5-fold enhancement in the emission intensity of $X^\text{-}$. We correlated the strain-induced broadening of both $X^0$ and $X^\text{-}$ emission linewidths with their deformation potentials to conclude that the change in $X^\text{-}$ binding energy is mainly governed by electron-phonon coupling mechanism. In this regard, we used a simple and cost-effective approach of imparting local strain on ML-WS$_2$ using spherical and shape-modified (see Methods) dielectric nanoparticles (NPs) as local stressors. The size distribution of NPs, together with the different levels of conformality of the flake at NP locations allowed us to explore strain in the range 0.1\,-\,2.0\,\%. We performed micro-photoluminescence (\textmu-PL) spectroscopy to investigate excitonic emissions and micro-Raman (\textmu-Raman) spectroscopy for quantifying and verifying strain of strained ML-WS$_2$.
	\section{Results}
	\par We started our investigation with \textmu-PL measurements on Sample\,1a. Figure\,\textcolor{blue}{\ref{Fig:strain1}(a)} shows the PL peak intensity map of Sample\,1a, displaying a strong light emission at all NP locations in the ML region of the flake due to a strain-induced funneling effect {\textcolor{blue}{\cite{moon2020dynamic,branny17}}. A one-to-one correlation of the strong light emission spots in the PL map with the dark contrast spots in the optical image can be seen in Fig. S1 of the Supplementary Information. The deposition of a flake on top of the NPs-coated substrate has resulted in formations of different levels of conformalities: from a full conformality (see an SEM image in the left-inset of Fig.\,\textcolor{blue}{\ref{Fig:strain1}(a)}) to a tent-like structure (see an SEM image in the right-inset of Fig.\,\textcolor{blue}{\ref{Fig:strain1}(a)}). The magnitude of local strain in the ML-WS$_2$ strongly depends on its extent of bending/conformality to the NPs and their size distribution, leading to a variation of local strain. This explicitly explains the different brightness of those NP-associated PL hotspots in Fig.\,\textcolor{blue}{\ref{Fig:strain1}(a)}. Other features of the PL signal have also been affected greatly by this variation in the local strain.
	\begin{figure*}[htb]
		\includegraphics{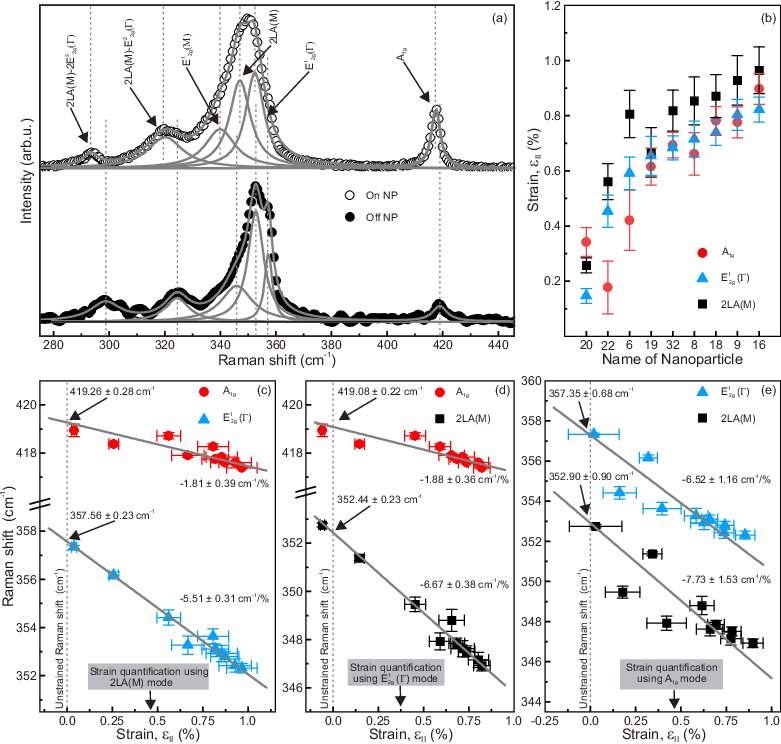}
		\caption{\textbf{Strain quantification and its verification via different Raman modes in Sample\,1a}: (a) µ-Raman spectra taken on-the-NP location (open circles, top) and unstrained region (closed circles, bottom) of ML-WS$_2$. Solid thick lines are the fits for the spectrum on-the-NP and unstrained locations. (b) Plot showing the distribution of strain on different NP locations. (c) Raman shift of E${^1_{2g}}$($\Gamma$) and A$_{1g}$ modes as a function of strain quantified from 2LA(M) mode. (d) Raman shift of 2LA(M) and A$_{1g}$ modes as a function of strain quantified from E${^1_{2g}}$($\Gamma$) mode. (e) Raman shift of E${^1_{2g}}$($\Gamma$) and 2LA(M) modes as a function of strain quantified from A$_{1g}$ mode. The grey solid lines represent the linear fits.}
		\label{Fig:strain2}	
	\end{figure*}
		
	\par In the next, we examine the effects of variation in local strain on emission energies of two important quasiparticles, $X^{0}$ and $X^\text{-}$, by acquiring PL spectra on all the NP locations as well as the unstrained regions in ML-WS$_2$. The statistical investigation of $E_{X^0}$, $E_{X^-}$, and $E_{b, \,X^-}$ for both the strained and unstrained ML-WS$_2$ is presented in Fig.\,\textcolor{blue}{\ref{Fig:strain1}(c)} in the form of Box-charts.
	Figure\,\textcolor{blue}{\ref{Fig:strain1}(b)} compares a PL spectrum (open circles) taken at the NP16 location from the ML-WS$_2$ region, representing one of the highly strained conditions with another PL spectrum (closed circles) from the unstrained ML-WS$_2$. We employed a two-peak fitting function (solid lines), a Gaussian function for high-energy $X^{0}$ emission peak (dotted lines) and a Lorentzian function for low-energy $X^\text{-}$ emission peak (dashed lines) for extracting the $X^0$ and $X^\text{-}$ emission energies. On the unstrained region, we obtained an $E_{b, \,X^-}$ of 32.3 meV which is well within the range (20\,-\,41\,meV) of values reported in literature \textcolor{blue}{\cite{mak2013tightly,ross2013electrical,mitioglu2013optical,michail2023tuning,he2016strain}}. Narrower distributions of $E_{X^0}$ and $E_{X^-}$, shown by solid boxes in the Box-charts are for unstrained ML-WS$_2$ and they indicate homogeneity of flake throughout the substrate. However, we observe wider distributions of $E_{X^0}$ and $E_{X^-}$ due to variation of strain at NP locations. On NP16, we observed a significant redshift of 36.8 meV in the $E_{X^0}$ and a relatively large red-shift of 48.8 meV in $E_{X^-}$ resulting a substantial increase in an $E_{b, \,X^-}$ of 15.5 meV as shown in Fig.\,\textcolor{blue}{\ref{Fig:strain1}(b)}. We obtain an average PL emission energy value of 2.008 eV (1.976 eV) (represented by small squares in solid Box-charts) corresponding to $E_{X^0}\left(E_{X^-}\right)$ for the unstrained regions.  As per literature, for unstrained ML-WS$_2$, the PL peak emission energies of $X^0$ and $X^\text{-}$ at room temperature are 2.018 eV and 1.975 eV \textcolor{blue}{\cite{plechinger2015identification}}. The values of $X^0$ and $X^\text{-}$ emission energies in the unstrained region of Sample\,1a are 2.009 eV and 1.976 eV. Due to the deposition of 75\,nm-thick SiO$_2$ film with a lower dielectric constant than the 272 nm-thick thermally-grown SiO$_2$ layer on the Si-substrate, we observed a lower $E_{X^0}$ in the unstrained region of ML-WS$_2$ \textcolor{blue}{\cite{hsu2019dielectric}}. On the NP locations, an average emission energy value of 1.979 eV (1.935 eV) (represented by small squares in open Box-charts) corresponding to $E_{X^0}\left(E_{X^-}\right)$ is observed. Analogous to the results on NP16, the average $E_{b, \,X^-}$ for the unstrained flake comes out to be 32.5 meV while the average $E_{b, \,X^-}$ from all measured NP locations is 44.3 meV which is giving an average increased $E_{b, \,X^-}$ of 11.8 meV as shown in the Fig.\,\textcolor{blue}{\ref{Fig:strain1}(c)}. The red-shifts of $E_{X^0}$ and $E_{X^-}$ indicate towards the significant amount of local tensile strain created at the NP locations due to the bending of the flake \textcolor{blue}{\cite{castellanos2013local}}.
	
	\par For a quantitative understanding of local strain, we performed \textmu-Raman Spectroscopy, in back-reflection geometry, on all investigated NP locations of ML-WS$_2$. Figure\,\textcolor{blue}{\ref{Fig:strain2}(a)} shows a comparison of Raman spectra of ML-WS$_2$ taken on the NP16 location (open circles) with the Raman spectra taken at unstrained region (closed circles) in Sample\,1a. We employed the Lorentzian peak fitting function to identify and extract the frequencies of all six different Raman modes in the Raman spectra of ML-WS$_2$ taken in the unstrained region and at the NP locations. In accordance with our expectations, we observed different Raman shifts for all the six Raman modes of ML-WS$_2$ as shown by dotted vertical lines in Fig.\,\textcolor{blue}{\ref{Fig:strain2}(a)}. We utilized three dominant and extensively studied Raman peaks: A$_{1g}$, E${^1_{2g}}$($\Gamma$) and 2LA(M) in the Raman spectra of ML-WS$_2$ as three different probes to calculate strain created in the ML-WS$_2$ due to underneath NPs. As these three different peaks correspond to different modes of atomic vibrations in the crystal lattice, they exhibit different peak shifts in response to the particular magnitude of strain. We quantified the magnitude of local strain on multiple NP locations in Sample\,1a, as shown in Fig.\,\textcolor{blue}{\ref{Fig:strain2}(b)} using the shift rate of -1.8 cm$^{-1}$, -5.7 cm$^{-1}$ and -6.3 cm$^{-1}$ /\% strain and unstrained Raman shift frequencies of 419, 357 and 353 cm$^{-1}$ for A$_{1g}$, E${^1_{2g}}$($\Gamma$) and 2LA(M) modes of vibration \textcolor{blue}{\cite{michail2023tuning,zhang2015phonon}}. The distribution of NP sizes together with different levels of conformality of flake on the NP locations (see figure S2 of Supplementary  Information) serve as local stress-inducing agents, enabling us to explore strain in the range of 0.1\,-\,1.0 \% in Sample\,1a.
	\begin{figure}[t]
		\includegraphics{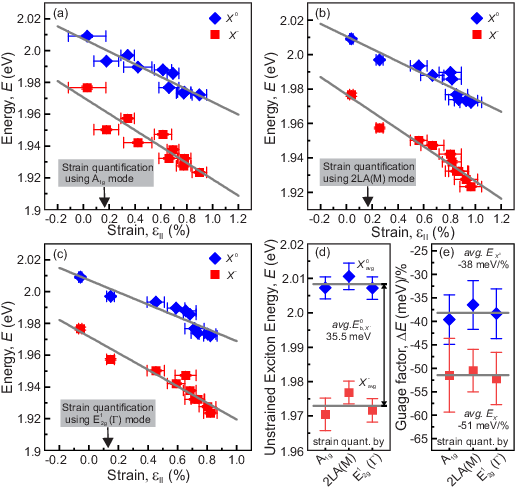}
		\caption{\textbf{Binding-energy tuning-rate per \% biaxial strain (gauge factor) of $X^\text{-}$ emission in ML-WS$_2$ in Sample\,1a}: $X^0$ and $X^\text{-}$ emission energies as a function of strain quantified using (a) A$_{1g}$, (b) 2LA(M) and (c) E${^1_{2g}}$($\Gamma$) Raman modes, showing unequal energy tuning rates leading to a change in the binding energy of $X^\text{-}$. (d) $X^0$ and $X^\text{-}$ emission energies of unstrained ML-WS$_2$ and (e) $X^0$ and $X^\text{-}$ energy gauge factors of ML-WS$_2$ quantified using A$_{1g}$, 2LA(M) and E${^1_{2g}}$($\Gamma$) Raman modes. The diamonds and squares are for $X^0$ and $X^\text{-}$, respectively. The solid lines in (a), (b) and (c) represent the linear fits and in (d) and (e) represent the averages.}
		\label{Fig:strain3}
	\end{figure}
	
	\par We further cross-verified the quantified strain through the following procedures. Figure\,\textcolor{blue}{\ref{Fig:strain2}(c)} show the Raman shifts of  E${^1_{2g}}$($\Gamma$) and A$_{1g}$ modes as a function of strain quantified from 2LA(M) mode. Similarly, Fig.\textcolor{blue}{\,\ref{Fig:strain2}(d)} shows the Raman shifts of 2LA(M) and A$_{1g}$  as a function of strain quantified from E${^1_{2g}}$($\Gamma$) mode and Fig.\,\textcolor{blue}{\ref{Fig:strain2}(e)} shows the Raman shifts of E${^1_{2g}}$($\Gamma$) and 2LA(M) as a function of strain quantified from A$_{1g}$ mode; the solid lines are the linear fits in all these three figures. It can be seen that the fitted values of Raman shifts of unstrained ML-WS$_2$ and shift rates of all three Raman modes of ML-WS$_2$ are in good agreement with the values reported in the literature (see also Table S1 of the Supplementary  Information). The slight mismatch in the shift rate of A$_{1g}$ peak from the reported shift rate can be attributed to the fact that A$_{1g}$ is sensitive to doping concentration in the material \textcolor{blue}{\cite{sasaki2016growth,dhakal2014confocal}}.
	
	\par Post acquiring the strain from Raman modes, we correlated the $E_{X^0}$ and $E_{X^-}$ in PL spectra of ML-WS$_2$ with the measured strain in Sample\,1a which is summarized in Fig.\,\textcolor{blue}{\ref{Fig:strain3}}. The local strains plotted in Fig.\,\textcolor{blue}{\ref{Fig:strain3}(a)}, \textcolor{blue}{(b)}, and \textcolor{blue}{(c)}, have been quantified using A$_{1g}$, 2LA(M) and E${^1_{2g}}$($\Gamma$) Raman modes, respectively. Taking into consideration that a small range of strains is accessed in this experiment, we performed linear fittings of these energy variations \textcolor{blue}{\cite{he2016strain}} and obtained values of $E_{X^0}^{0} \left(E_{X^-}^{0}\right)$, emission energy of $X^{0} \left(X^\text{-}\right)$ transition in an unstrained ML-WS$_2$, that are plotted in Fig.\,\textcolor{blue}{\ref{Fig:strain3}(d)}. From these fits (solid lines in Fig.\,\textcolor{blue}{\ref{Fig:strain3}}), we also obtained $\Delta E_{X^0}/\% \left(\Delta E_{X^-}/\%\right)$ strain, an energy-gauge-factor of $X^0$ ($X^\text{-}$) transition in strained ML-WS$_2$, that are plotted in Fig.\,\textcolor{blue}{\ref{Fig:strain3}(e)}. As can be seen in Fig.\,\textcolor{blue}{\ref{Fig:strain3}(d)}, we obtained slightly different values of $E_{X^0}$ and $E_{X^-}$ for unstrained ML-WS$_2$ using different strain estimating Raman modes. The average (solid line) of these three $E_{X^0}$($E_{X^-}$) values is 2.008 (1.973) eV, which is similar to the value obtained in the statistical PL investigation of unstrained ML-WS$_2$. Similarly, we obtained the average $\Delta{E_{X^0}}/\%\left(\Delta{E_{X^-}}/\%\right)$ strain of -38\,$\pm$\,3.0 (-51\,$\pm$\,3.5)\,meV/\% strain for emission energies of $X^0$ ($X^\text{-}$) transition which are well within the range of -11 to -130  meV /\% strain, reported in the literature \textcolor{blue}{\cite{wang2015strain,michail2023tuning}}. As the emission of light due to $X^\text{-}$ happens at a lower energy than $X^0$, the difference in $X^0$ and $X^\text{-}$ gauge factors lead to a remarkable increase in the binding energy of $X^\text{-}$ with a tuning rate, \(\Delta\)$E_{b, \,X^-}$/\%, of 13\,$\pm$\,4.5\,meV/\% strain for Sample\,1a.
	\begin{figure}[htb]
		\includegraphics{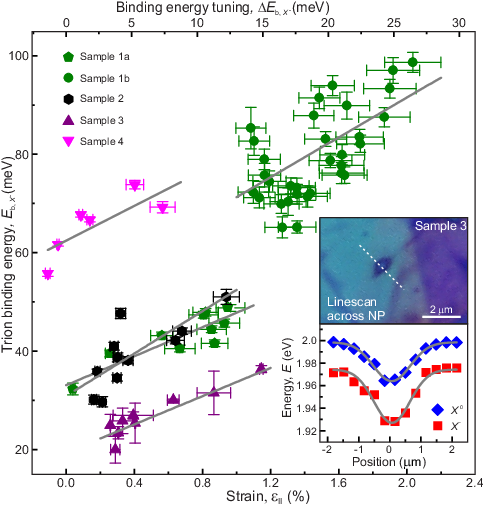}
		\caption{\textbf{Strain tuning-rate of trion's binding energy, \(\Delta\)$E_{b, \,X^-}$/\% across vraious samples }: Strain-dependent trion's binding energy, $E_{b, \,X^-}$ for Sample\,1a (pentagons), Sample\,1b (circles), Sample\,2 (hexagons), Sample\,3 (triangles) and Sample\,4 (inverted triangles), showing a very similar tuning-rates of trion binding energy per \% of biaxial strain. Inset:  The optical micrograph of Sample\,3. A white dotted line is representing the PL linescan (top) and the change in $X^0$ (diamonds) and $X^\text{-}$ (squares) PL peak energy positions as a function of distance across the NP location (bottom).}
		\label{Fig:strain4}
	\end{figure}
	
	\par  We attribute this significant change in $E_{b, \,X^-}$ to the pronounced variation in the local strain that can be qualitatively understood in terms of strain-induced variations in the interaction energies between the constituent particles of $X^0$ and $X^\text{-}$ in ML-WS$_2$. The quasiparticle $X^\text{-}$ consists of 2 electrons ($e^-$) and one hole ($h$), and the interaction energies between these constituent particles play a significant role in contributing to the $E_{b, \,X^-}$ of ML-WS$_2$. Biaxial strain induces changes in both electron-hole interaction energy ($J_{eh}$) as well as electron-electron interaction energy ($J_{ee}$). However, for a strongly 2D-confined system, under the assumption that strain induces negligible changes in the single-particle confinement energies, the rate of change of $E_{b, \,X^-}$ per unit strain is given by:
	\begin{eqnarray}\label{eq:change in Trion binding energy}
		\frac{d}{d\epsilon_{\parallel}}\left({E_{b,X^-}}\right)&\approx& \frac{d}{d\epsilon_{\parallel}}\left({J}_{eh}\right) - \frac{d}{d\epsilon_{\parallel}}\left({J}_{ee}\right)
	\end{eqnarray}
	Ding et al., see Ref.\,[\textcolor{blue}{\onlinecite{ding2010tuning}}] have shown that with increasing biaxial tensile strain, the decrease in the second term in Eq.\,\textcolor{blue}{(\ref{eq:change in Trion binding energy})} is more significant as compared to the first term, therefore rendering a positive value of d$E_{b, \,X^-}$/d$\epsilon_{\parallel}$. It clearly indicates the increase in $X^\text{-}$ binding energy for ML-WS$_2$ via the introduction of biaxial tensile strain.
	\par To authenticate this large tuning rate, \(\Delta\)$E_{b, \,X^-}$/\%, we investigated four more samples with optically active material ML-WS$_2$ and findings are summarized in Fig.\,\ref{Fig:strain4}. It is expected that strain will decrease as we move away from the centre of NPs. Therefore, in Sample\,3, a single and isolated NP location was investigated by performing a line-scan PL measurement. The optical micrograph of the Sample\,3 showing the isolated NP location and changes in the $X^0$ and $X^\text{-}$ energies are shown in the inset of Fig.\,\textcolor{blue}{\ref{Fig:strain4}}. We utilized a gauge factor \(\Delta\)$E_{X^0}$/\% strain, -38 meV/\% strain that is obtained in Sample\,1a, to estimate strain in all the samples. Figure\,\textcolor{blue}{\ref{Fig:strain4}} shows the variation of $E_{b, \,X^-}$ as a function of strain for all investigated samples where the pentagons, circles, hexagons, triangles and inverted triangles represent the measured data points for Sample\,1a, Sample\,1b, Sample\,2, Sample\,3 and Sample\,4, respectively. \(\Delta\)$E_{b, \,X^-}$/\% obtained from the linear fitting of measured data from all five samples are summarized in Table\textcolor{blue}{\,\ref{tab:B.E.}}, and it shows that \(\Delta\)$E_{b, \,X^-}$/\% strain for all samples are similar and within the error bars. Conclusively, we state that on average, we obtain the tuning rate of $X^\text{-}$ binding energy, \(\Delta\)$E_{b, \,X^-}$/\% strain of 17.5\,$\pm$\,2.5\,meV/\% strain.
	\begin{figure*}[htb]
		\includegraphics{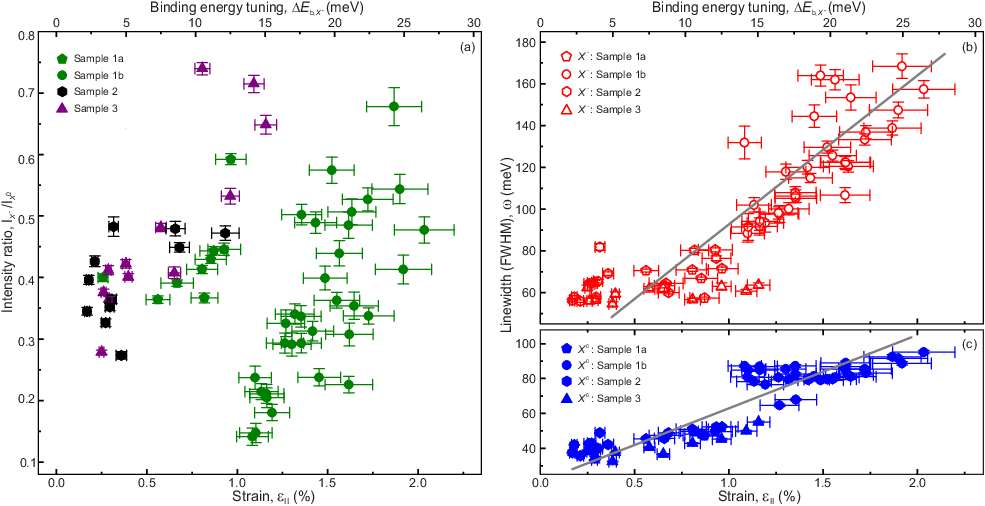}
		\caption{\textbf{Strain-dependent PL emission intensity and linewidth (FWHM) of $X^0$ and $X^\text{-}$ emission across different samples:} (a) Strain-dependent ratio of the intensity of $X^\text{-}$ emission peak ($I_{X^-}$) to the intensity of $X^0$ emission peak for Sample\,1a (pentagons), Sample\,1b (circles), Sample\,2 (hexagons) and Sample\,3 (triangles). Strain-dependent linewidth of (b) $X^\text{-}$ and (c) $X^0$ emission peaks, for all the investigated samples.  Pentagons, circles, hexagons and triangles in (b) and (c) are for Sample\,1a, Sample\,1b, Sample\,2 and Sample\,3, respectively.}
		\label{Fig:strain5}
	\end{figure*}
	\begin{table}[htb]
		\fontsize{9.5pt}{9.5pt}\selectfont
		\renewcommand{\arraystretch}{1.6}
		\caption{$X^\text{-}$ Binding-energy tuning-rate per \% biaxial strain (gauge factor) of ML-WS$_2$ for all samples}
		\label{tab:B.E.}
		\begin{center}
			\begin{tabular}{|p{2cm}|p{5cm}|}
				\hline
				\multirow{2.7}{2.7cm}{\hspace{0.6cm}Sample} &  Tuning-rate of trion binding energy, \(\Delta\)$E_{b, \,X^-}$/\% strain\\
				& \hspace{2cm} meV/\%\\
				\hline
				\hspace{0.4cm}Sample\,1a & \hspace{1.7cm} 13.0 $\pm$ 4.5  \\
				\hspace{0.4cm}Sample\,1b & \hspace{1.7cm} 20.0 $\pm$ 5.0 \\
				\hspace{0.4cm}Sample\,2  & \hspace{1.7cm} 21.5 $\pm$ 5.5  \\
				\hspace{0.4cm}Sample\,3  & \hspace{1.7cm} 14.5 $\pm$ 2.5  \\
				\hspace{0.4cm}Sample\,4  & \hspace{1.7cm} 17.5 $\pm$ 7.0  \\ \hline
		\end{tabular}\end{center}
	\end{table}
	
	\par Looking at the fact that the local strain has increased the energetic separation between $X^0$ and $X^\text{-}$ transitions, it is expected to alter the transition probability of both transitions. In the following, we inspect this aspect by monitoring the emission intensity and linewidth of both the transitions, and the outcome is summarized in Fig.\,\textcolor{blue}{\ref{Fig:strain5}}. Figure\,\textcolor{blue}{\ref{Fig:strain5}(a)} shows the  $I_{X^-}$/$I_{X^0}$ ratios for all the investigated samples, where $I_{X^0}$($I_{X^-}$) is the intensity of $X^0$($X^\text{-}$) that are plotted as a function of strain. In each investigated sample, we noticed a clear increase in the $I_{X^-}$/$I_{X^0}$ ratio with increasing strain. Along with the increase in the $I_{X^-}$/$I_{X^0}$, we observed two different sets of data points; Although both the data sets show a similar increase in the intensity ratio $I_{X^-}$/$I_{X^0}$, the minimum intensity ratio of one data set is higher than the other because of different dielectric environment and sample preparation conditions. However, irrespective of the value of strain quantified for different samples, we observed up to $\approx$ $5 \times$ enhancement in the $I_{X^-}$/$I_{X^0}$ ratio on the introduction of up to 2\% tensile strain. This observation is in accordance with the mass action law exhibiting an increase in the $I_{X^-}$/$I_{X^0}$ with increasing $E_{b, \,X^-}$ that is increasing with strain \textcolor{blue}{\cite{kesarwani2022control,dastidar2024optically}}. Also, we monitored the strain-induced broadening (increase) in the full width at half maximum\,(FWHM), linewidth, of $X^0$ and $X^\text{-}$ emission peaks for all investigated samples, which is summarized in Fig.\,\textcolor{blue}{\ref{Fig:strain5}}. As per literature, the broadening in the linewidth of excitonic emission peak in the PL spectra as a function of strain has only been observed for MoS$_2$ with a broadening rate of 10\,-\,15 meV/\% \textcolor{blue}{\cite{niehues2018strain,khatibi2018impact}} while MoSe$_2$, WS$_2$ and WSe$_2$ showed narrowing (decrease) in linewidth as function of strain with rates 5, 8 and 20 meV/\%, respectively \textcolor{blue}{\cite{khatibi2018impact}}. Our linewidth vs strain data for both $X^0$ and $X^\text{-}$ peaks of ML-WS$_2$ plotted in Fig.\,\textcolor{blue}{\ref{Fig:strain5}(b),\,(c)} shows a clear broadening in the linewidth under tensile strain. The linear fits of the measured data produced a linewidth shift rate of 42 meV/\% (71 meV/\%) strain for $X^0\left(X^\text{-}\right)$ which shows that the phonon is interacting more strongly with $X^\text{-}$ than $X^0$. Shen et al., see Ref.\,[\textcolor{blue}{\onlinecite{shen2016strain}}], have shown that the valence bands of WSe$_2$ are not much affected by strain up to 2\%. Thus, we further investigated these strain effects on excitons through the coupling strength of electron-phonon interaction that is given by:
	
	\begin{equation}
		\beta \propto \frac{DP^2}{Y}
	\end{equation}
	
	where, DP is the gap deformation potential of the quasiparticle, and Y is Young's Modulus of elasticity. In this work, since we are monitoring the $\Delta{E_{X^0(X^-)}}$, we can calculate the DP of both the transitions given by \textcolor{blue}{\cite{indrajeet2024}}:
	
	\begin{equation}
		DP_{X^0(X^-)} = \frac{\Delta{E_{X^0(X^-)}}}{(2 - \rho) \cdot \epsilon_{\parallel}}
	\end{equation}
	
	where, $\rho$ (= 0.19) is the Poisson's ratio of ML-WS$_2$, $\epsilon_{\parallel}$ refers to local strain, which we have also measured in this work. Since, $\Delta{E_{X^0(X^-)}}$ per unit strain can be obtained from Fig.\,\textcolor{blue}{\ref{Fig:strain3}}, the calculated $DP_{X^0(X^-)}$ is 2.10\,(2.82) eV, giving rise to coupling strengths ratio of electron-phonon interaction for $X^0$ and $X^\text{-}$, $\beta_{X^-}$/$\beta_{X^0}$ = 1.80 $\pm$ 0.27. It is striking to see that aforementioned ratio of coupling strengths of electron-phonon interaction for $X^0$ and $X^\text{-}$, which is obtained from the strain-induced energy changes in them, also agrees well with the ratio of linewidth shift rates of $X^0$ and $X^\text{-}$, $\Delta \omega_{X^{-}}$/$\Delta \omega_{X^{0}}$ per percent strain, rendering a value of 1.71 $\pm$ 0.14. From this, we conclude that the strain-induced changes in the binding energies of $X^\text{-}$ can be explained by electron-phonon interactions.

	\section {Discussions}
	In this work, we report the large increase of $X^\text{-}$ binding energy by 34 meV due to strain-enhanced electron-phonon coupling in ML-WS$_2$. To demonstrate this strain-induced variation, we investigated the emission energies of $X^0$ and $X^\text{-}$ by performing the \textmu-PL measurements on unstrained and NP locations in ML-WS$_2$ across different samples. The strain in ML-WS$_2$ was quantified by performing the Raman measurements as described in Fig.\,\textcolor{blue}{\ref{Fig:strain2}(a)}\,and\,\textcolor{blue}{(b)}. Three dominant and extensively studied Raman modes: A$_{1g}$, E${^1_{2g}}$($\Gamma$) and 2LA(M) modes were employed for this strain quantification. The cross-verification of strain was done by plotting the Raman shift frequencies of two Raman modes against the strain quantified from the third Raman mode. This process was utilized for all three Raman modes as described in Fig.\,\textcolor{blue}{\ref{Fig:strain2}}. We further correlated the strain-induced variations in the peak emission energies of $X^0$ and $X^\text{-}$ transitions of ML-WS$_2$ with the quantified strain as summarized in Fig.\,\textcolor{blue}{\ref{Fig:strain3}} and obtained the energy gauge factor of $X^{0} \left(X^\text{-}\right)$, $\Delta{E_{X^0}}/\%\left(\Delta{E_{X^-}}/\%\right)$ strain, of -38\,$\pm$\,3.0 (-51\,$\pm$\,3.5)\,meV/\% biaxial strain leading to $X^\text{-}$ binding energy tuning rate,\,\(\Delta\)$E_{b, \,X^-}$/\% strain, of 13\,$\pm$\,4.5\,meV/\% strain as shown in Fig.\,\textcolor{blue}{\ref{Fig:strain3}(e)}. To further verify this strain-induced large tuning rate, we investigated the PL spectra of strained ML-WS$_2$ from four additional samples as shown in Fig.\,\textcolor{blue}{\ref{Fig:strain4}}, where the strain was quantified using the gauge factor of -38 meV/\% obtained from Sample\,1a. The distribution in sizes of NPs and different levels of conformality of the flake on NP locations enabled us to explore strain in the range 0.1\,-\,2.0\,\% across all samples. Utilizing the excitonic gauge factor, where strain was quantified via Raman spectroscopy of Sample\,1a, we achieved similar $X^\text{-}$ binding-energy tuning rates for other four samples, highlighting the consistency of this approach. Conclusively, we obtained an average \(\Delta\)$E_{b, \,X^-}$/\% strain of 17.5$\pm$ 2.5\,meV/\% for all the investigated samples irrespective of the surrounding dielectric environment of ML-WS$_2$ and the sample preparation conditions summarized in Table\textcolor{blue}{\,\ref{tab:B.E.}} and we have shown that this large change in binding energy can be understood in terms of the interaction energies between the constituent particles in $X^0$ and $X^\text{-}$ quasiparticles. Alongside the large variation in $X^\text{-}$ binding energy, we also investigated the strain-induced linewidth broadening shown in Fig.\,\textcolor{blue}{\ref{Fig:strain5}(b)}\,and\,\textcolor{blue}{(c)} and the deformation potentials of both the $X^0$ and $X^\text{-}$ transitions and hence, elucidated the significant role of strain-enhanced electron-phonon coupling in increasing the $X^\text{-}$ binding energy. This strain-induced increase in the $E_{b, \,X^-}$ followed by enhanced stability and efficient formation of $X^\text{-}$ holds relevance for future $X^\text{-}$ based nano-opto-electro-mechanical systems even at elevated temperatures.
	
	\section*{Methods}
	\subsection*{Sample preparation}
	The studied samples consist of three different layered structures. Samples 1\,\&\,4 (Sample\,2) consist of mechanically exfoliated monolayer WS$_2$ deposited on shape-modified (spherical) NPs distributed over a SiO$_2$ (272 nm) / Si substrate. Utilizing the spin-coating method, we obtained a random distribution of individual NPs that are well isolated from each others on all samples. The size of spherical NPs (Sigma-Aldrich) falls typically in the range of 125-175\,nm. These NPs act as nano stressors, creating local strain in ML-WS$_2$. NP concentration in ethanol was optimized to prevent agglomeration. In Sample\,1\,\&\,4, the process of shape modification of NPs was accomplished through deposition of 75\,nm-thick SiO$_2$ film on spherical-NPs distributed over a SiO$_2$/Si-substrate using an e-beam Physical Vapour Deposition technique. The active material, ML-WS$_2$, is then deposited on top of shape-modified-NPs. In Sample\,3, NPs were spin-coated onto the bottom-hBN deposited on SiO$_2$/Si-substrate, followed by transfer of ML-WS$_2$, which is further encapsulated by a thin-hBN flake. Both the WS$_2$ and hBN flakes were mechanically exfoliated from their bulk counterparts. The ML-WS$_2$ flakes were qualitatively identified by optical color contrast and further confirmed by PL \& Raman spectroscopy. We employed the conventional dry-transfer method using polydimethylsiloxane (PDMS) stamps to transfer the flakes on top of each other \textcolor{blue}{\cite{castellanos14alldry}}. We also noticed wrinkles formation during the top-thin hBN transfer because of the high elastic modulus of the hBN \textcolor{blue}{\cite{falin2017mechanical}} compared to ML-WS$_2$ \textcolor{blue}{\cite{liu2014elastic}}. As a standard protocol of sample processing, we employed vacuum annealing at 200$^{\circ}$C for 5 hours, followed by natural cooling in a rapid-thermal-annealing (RTA) system for Sample\,1 \& 3 \textcolor{blue}{\cite{surendran2024nanoparticle}}.
	\subsection*{PL and Raman Spectroscopy}
	Photoluminescence and Raman signals from all three samples were collected using a home-built confocal microscopy setup equipped with a microscope objective (NA\,=\,0.75), yielding diffraction-limited spatial resolutions. The samples were mounted on an XY scanner combined with a $\pm$ 2.5 mm XYZ nanopositioner stack (Attocube). For generating the space maps of PL signals, the samples were moved using the scanner providing a scanning range of 50\,\textmu m. A diode-pumped solid-state CW laser emitting at $\lambda$\,=\,532\,nm was used as an exciting source for both PL and Raman measurements. An ultra-steep long pass filter designed at an edge of 533.3\,nm (Semrock) was used to suppress the laser light from entering into the spectrometer. All the spectra were acquired with a 0.5\,m focal length spectrometer combined with a water-cooled charge-coupled device camera providing a best spectral resolution of $\sim$125\,\textmu eV at $\lambda$\,=\,532\,nm on an 1800 lines/mm grating. A grating with 150 lines/mm giving a spectral resolution of 2.5 meV at $\lambda$\,=\,532\,nm was used for acquiring large wavelength range spectra.
	
	\begin{acknowledgments}
	We thank E.S. Kannan, S. R. Parne, A. Rahman, and S. J. Cheng for the fruitful discussion and A. Rastelli for the data analysis software. This work was supported by the DST Nano Mission grant (DST/NM/TUE/QM-2/2019) and the matching grant from IIT Goa. I.D.P. thanks The Council of Scientific $\&$ Industrial Research (CSIR), New Delhi, for the doctoral fellowship. K.W. and T.T. acknowledge support from the JSPS KAKENHI (Grant Numbers 21H05233 and 23H02052) and World Premier International Research Center Initiative (WPI), MEXT, Japan.
	\end{acknowledgments}	
	
    \section*{Supporting Information}
    \setcounter{figure}{0}
    \renewcommand{\thefigure}{S\arabic{figure}}
    \setcounter{table}{0}
    \renewcommand{\thetable}{S\arabic{table}}
    \begin{table*}[htb]
    	\fontsize{9pt}{9pt}\selectfont
    	\renewcommand{\arraystretch}{1.4}
    	\caption{Raman shifts of A$_{1g}$, 2LA(M) and E${^1_{2g}}$($\Gamma$) modes in unstrained ML-WS$_2$ and changes in the Raman modes per \% biaxial strain: Literature \& this work}
    	\label{tab:Raman}
    	\begin{center}
    		\begin{tabular}{|p{3cm}|p{2.5cm}|p{2.9cm}|p{2.9cm}|p{1.8cm}|p{1.8cm}|p{1.8cm}|}	
    			\hline
    			\multirow{3}{3.2cm}{Raman mode used for strain quantification} & \multicolumn{2}{c|}{From Literature} & Raman shift of unstrained ML-WS$_2$ in this work &\multicolumn{3}{c|}{Changes in the Raman shift per \% biaxial strain} \\ \cline{2-3}
    			& Raman shift of unstrained ML-WS$_2$. Theory: (Ref.\,\onlinecite{zhang2015phonon}) & Changes in Raman shift per \% biaxial strain. Experiment: (Ref.\,\onlinecite{michail2023tuning}) & &\multicolumn{3}{c|}{}\\
    			& \hspace{1.10cm}(cm$^{-1}$) & \hspace{0.65cm}(cm$^{-1}$/\%)& \hspace{1.2cm}(cm$^{-1}$) &\multicolumn{3}{c|}{(cm$^{-1}$/\%)} \\
    			\cline{5-7}
    			&  & & &\hspace{0.6cm} A$_{1g}$ &\hspace{0.5cm}E${^1_{2g}}$($\Gamma$) &\hspace{0.25cm} 2LA(M)\\ \hline
    			\hspace{1.15cm} A$_{1g}$ &\hspace{1.0cm} 419 & \hspace{0.7cm}-1.8 $\pm$ 0.1 & \hspace{0.4cm} 418.94 $\pm$ 0.26 &\hspace{0.7cm} -- & \hspace{0.1cm}-6.52 $\pm$ 1.16 & -7.73 $\pm$ 1.53  \\
    			\hspace{1.0cm} E${^1_{2g}}$($\Gamma$)& \hspace{1.07cm}357 & \hspace{0.7cm}-5.7 $\pm$ 0.2 & \hspace{0.45cm}357.33 $\pm$ 0.06 & -1.88 $\pm$ 0.36  &\hspace{0.8cm}-- & -6.67 $\pm$ 0.38 \\
    			\hspace{0.9cm} 2LA(M) &\hspace{1.0cm} 353 & \hspace{0.7cm}-6.3 $\pm$ 0.5 & \hspace{0.45cm}352.75 $\pm$ 0.08 &-1.81 $\pm$ 0.39  & \hspace{0.1cm}-5.51 $\pm$ 0.31 &\hspace{0.7cm}-- \\ \hline
    		\end{tabular}
    	\end{center}
    \end{table*}

    \begin{figure}[h]
    	\includegraphics[width=8cm]{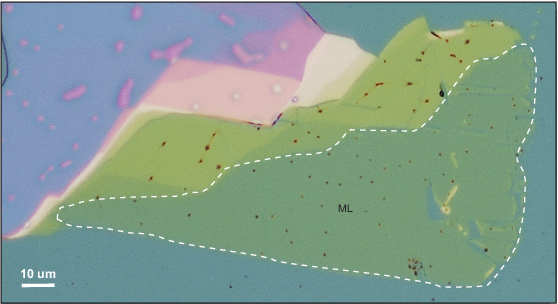}
    	\caption{\textbf{Optical micrograph showing monolayer (dotted line) and thicker regions of WS$_2$ flake in Sample\,1a}}
    	\label{Fig:supp_strain1}
    \end{figure}

    \begin{figure}[h]
    	\includegraphics[width=8cm]{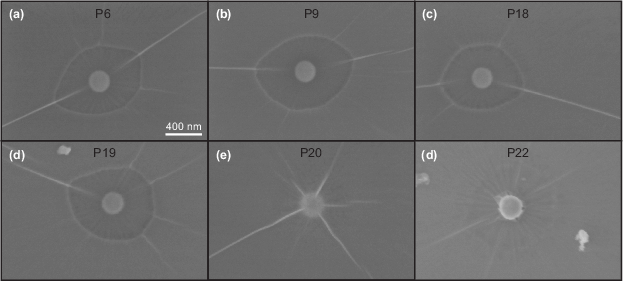}
    	\caption{\textbf{SEM images of NPs showing different flake conformalities according to NP surface in Sample\,1a}}
    	\label{Fig:supp_strain2}
    \end{figure}

    \section{Coulombic interactions in bound state quasiparticles}
    The emission energies of the bound state quasi-particles ($X^0$ \& $X^\text{-}$) depend on several factors including the constituent particle (electron and hole) energies, interaction energy between the particles and size and shape of the confinement potential, which is shown by Eq. 1.
    \begin{eqnarray}\label{eq:Exciton and Trion emission energy}
    	E_{X^0} &=& E_e + E_h + E_g - J_{eh}\nonumber\\
    	E_{X^-} &=& 2E_e + E_h + E_g - 2J_{eh} + J_{ee}
    \end{eqnarray}
    \par where, E$_e$ and E$_h$ represents the single particle energy of electron and hole respectively, E$_g$ represents the band-gap energy of ML-WS$_2$ and J$_{eh}$ and J$_{ee}$ are the Coulombic interaction energies of electron-hole and electron-electron interaction, respectively. As per the above equations, $E_{b,X^-}$ is given by :
    \begin{eqnarray}\label{eq:Trion binding energy}
    	E_{b,X^-} &=& E_{X^0} - E_{X^-} \nonumber\\
    	E_{b,X^-} &=& J_{eh} - J_{ee} - E_e
    \end{eqnarray}

    As seen in Eq.\,\textcolor{blue}{(\ref{eq:Trion binding energy})}, the binding energies of $X^0$ and $X^\text{-}$ are strongly influenced by the interaction energies between their constituent particles. 

    \par Figure\,\textcolor{blue}{\ref{Fig:supp_strain1}} shows an optical micrograph of a WS$_2$ flake deposited on top of the shape-modified-NPs for Sample\,1a. The presence of individual NPs beneath the flake, consisting of ML and thicker regions, noticeable via dark contrast spots in the image. These spots show one-to-one correlation with the strong light emission spots in the PL map of the flake.

\end{document}